\documentclass{iaus}
\usepackage{graphicx}

\title{Asymmetric warps in  disk galaxies: dependence on dark matter halo}

\author{K. Saha$^1$ \& C. J. Jog$^1$ }

\affiliation{$^1$Indian Institute of Science, Bangalore 560012, India; 
email: kanak@physics.iisc.ernet.in \break}

\pubyear{2006}
\volume{235}  
\date{?? and in revised form ??}
\jname{Proceedings Title IAU Symposium}
\editors{F.Combes \& J. Palous, eds.}
\begin{document}

\maketitle
\begin{abstract}
Recent observations have shown that most of the warps in the disk galaxies are 
asymmetric. However there exists no generic mechanism to generate these asymmetries in warps. We have shown that a rich variety of possible asymmetries in the z-distribution of the spiral galaxies can naturally arise due to a dynamical wave interference between the first two bending modes i.e. bowl-shaped mode(m=0) and S-shaped warping mode(m=1) in the galactic disk embedded in a dark matter halo.
We show that the asymmetric warps are more pronounced when the dark matter content within the optical disk is lower as in early-type galaxies.
\keywords{galaxies: kinematics and dynamics - galaxies: spiral - galaxies: structure }

\end{abstract}

\vspace{-5mm}
\firstsection 
\section{Generating asymmetric warps}
We study the dynamics of bending modes in a differentially rotating self-gravitating disk embedded in a dark matter halo. We show that the dynamical equation of a small bending with azimuthal wavenumber $m$ reduces to a compact quadratic eigenvalue problem when the galactic disk is approximated as a system of uniformly spaced rings ( see Saha \& Jog 2006 ). We notice that the bowl-shaped mode(m=0) oscillates much faster than the integral-sign warping mode (m=1) of the disk. We use the linear and time dependent superposition of the m=0 and m=1 mode to generate various dynamical asymmetries in warps including the extreme L-shaped or one-sided warps and asymmetric U-shaped warps. So in our picture the presence of bowl-shaped mode is the primary reason for the various asymmetric warps in the disk.       

\vspace{-6mm}
\section{Asymmetry-index and dark matter content in the disk galaxies } 
The asymmetry-index ($\alpha_{asym}$) is basically a quantitative measure of the asymmetric warps in the disk. $\alpha_{asym}$ varies from 0 (corresponding to symmetric warps) to 1.0 (corresponding to L-shaped warps). 

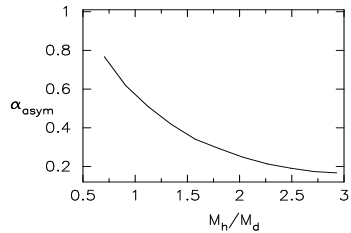
\begin{figure}[!h]
\begin{center}
{\rotatebox{270}{\resizebox{3.0cm}{4.5cm}{\includegraphics{kanaksaha_fig1.ps}}}}


\end{center}
\end{figure}
The figure shows that more the dark matter within the disk less is the value of $\alpha_{asym}$. We can draw a second conclusion that early type disks are more likely to show asymmetry. 

\begin{acknowledgments}
\noindent  K. Saha would like to thank IISc, DST \& CSIR, India and the XXVI IAU meeting grant
\end{acknowledgments}
\noindent{\bf Reference:}
\noindent Saha, K. \& Jog, C. J. 2006, A \& A, 446, 897

\end{document}